\documentclass[journal]{IEEEtran}
\usepackage{graphicx}
\usepackage{amsmath}
\usepackage{amsfonts}
\usepackage{amssymb}

\newtheorem{theorem}{Theorem}
\newtheorem{lemma}{Lemma}

\begin{document}

\title{\huge Multiple-access Fading Channel with Wireless \\ Power Transfer and Energy Harvesting}
\author{Zoran Hadzi-Velkov, Nikola Zlatanov, and Robert Schober 
\vspace{-3mm}
\thanks{Z. Hadzi-Velkov is with the Faculty of Electrical Engineering and Information Technologies, Ss. Cyril and Methodius University, Skopje, Macedonia (email: zoranhv@feit.ukim.edu.mk).}
\thanks{N. Zlatanov is with the Department of Electrical and Computer Engineering, University of British Columbia, Vancouver, Canada (email: zlatanov@ece.ubc.ca).}
\thanks{R. Schober is with the Institute of Digital Communications, Friedrich-Alexander University, Erlangen, Germany (email: schober@lnt.de).}
}

\markboth{accepted for journal publication}{Shell \MakeLowercase{\textit{et al.}}: Bare Demo of IEEEtran.cls for Journals} \maketitle

\begin{abstract}
We consider the achievable average rates of a multiple-access system, which consists of $N$ energy-harvesting users (EHUs) that transmit information over a block fading multiple-access channel (MAC) and a base station (BS) that broadcasts radio frequency (RF) energy to the EHUs for wireless power transfer. The information (over the uplink) and power (over the downlink) can be transmitted either in time division duplex or frequency division duplex. For the case when the EHUs' battery capacities and the number of transmission slots are both infinite, we determine the optimal power allocation for the BS and the optimal rates and power allocations for the EHUs that maximize the achievable rate region of the MAC. The resulting online solution is asymptotically optimal, and also applicable for a finite number of transmission slots and finite battery capacities.
\end{abstract}

\vspace{2mm}


\vspace{2mm}

\section{Introduction}
Energy harvesting (EH) technology may provide a perpetual power supply to energy-constrained wireless systems such as sensor networks. To maximize the system performance, the EH transmitters should adapt their output powers based on the energy harvested up to the time of transmission (causality constraint), whereas theoretical performance bounds are often determined for non-causally adapted output powers [\ref{litA1}]-[\ref{lit5}]. The EH from the environment (e.g., solar or wind) is an intermittent process, which can be mitigated by using wireless power transfer (WPT), based on far-field radio frequency (RF) radiation from a distant energy source. If the same signal is used for simultaneous energy and information transfer, a fundamental tradeoff exists between energy transfer and the achievable rate, as has been shown, e.g., for the noisy channel [\ref{litA3}], the multiple-input multiple-output (MIMO) broadcast channel [\ref{litA4}], and multiple-access channel (MAC) [\ref{litA5}].

In this paper, we study an EH network that consists of a base station (BS) and multiple EH users (EHUs), where the BS broadcasts RF energy to the EHUs over the downlink, and the EHUs send information simultaneously to the BS over the uplink MAC (denoted here as the EH MAC). Such a scenario is relevant, e.g., for sensor networks operating in hostile or inaccessible environments. Only a few existing works consider similar systems [\ref{litB2}], [\ref{litB3}], and only [\ref{litB3}] studies the achievable rates of the EHUs in block fading channels. In [\ref{litB3}], the EHUs send their information to the BS using time-division-multiple-access (TDMA), and, for each fading state, the EHUs' sum-rate is maximized by jointly optimizing the duration of the downlink interval allocated to the BS and the duration of the uplink intervals allocated to each of the EHUs.

The system model considered in this paper is more general: (a) The EHUs are allowed to send their information simultaneously to the BS, instead of via TDMA. (b) The EHUs are equipped with EH batteries that can store energy over multiple fading states (slots). (c) The BS transmit power may change from slot to slot subject to a long-term average power constraint. Assuming only availability of causal channel state information (CSI), we propose online rate and power allocation strategies for the EHUs and the BS so as to maximize the average achievable rates in the EH MAC.

\vspace{-0mm}

\section{System and Channel Model}
\vspace{0mm}
The network consists of one BS and $N$ EHUs. The time is divided into slots of equal duration, whose total number $M$ satisfies $M\to\infty$. The $n$th EHU's uplink and downlink channels are affected by independent channel fading and additive white Gaussian noise (AWGN) with power $N_0$. The fading in each channel is a stationary and ergodic random process, and follows the block fading model (i.e., they are constant in each slot but change from one slot to the next). In slot $i$, the fading power gains of the $n$th EHU's uplink and downlink channels are denoted by $x_n'(i)$ and $y_n'(i)$, respectively. For convenience, these gains are normalized by the AWGN power, such that $x_n(i) = x_n'(i)/N_0$ and $y_n(i) = y_n'(i)/N_0$. We assume that the BS and EHUs operate in half-duplex mode. Two multiplexing schemes for energy  and information transfer are possible:

(1) Time-division information and power transfer (TDT): Here, the uplink and downlink transmissions occur in the same frequency band but in different time slots. Thus, the $i$th time slot is dedicated to either uplink or downlink transmission. The uplink and downlink channels are assumed reciprocal, i.e., $x_n(i) = y_n(i)$. To mathematically model TDT,  we introduce a ``scheduling" variable $a_i$, defined  as
\begin{equation} \label{rav15}
a_i =\left\{
\begin{array}{rl}
1, & \text{slot $i$ is used for downlink transmission,}
\\
0, & \text{slot $i$ is used for uplink transmission}.
\end{array}
\right.
\end{equation}

(2) Frequency-division information and power transfer (FDT): Here, the uplink and downlink transmissions occur simultaneously but in two different frequency bands. Thus, in the $i$th time slot, simultaneous information and energy transmission is performed in the uplink  and  downlink, respectively. In this case, the power gains $x_n(i)$ and $y_n(i)$ are assumed to be independent. The downlink may require negligible spectrum resources (e.g., a "power carrier" sinusoid may be used).

The BS output power in the $i$th slot is $P_0(i)$. Two types of constraints are imposed on $P_0(i)$: (a) A peak power constraint, $P_0(i) \leq P_{max}$, and (b) an average power constraint, $E[a_i P_0(i)]  \leq P_{avg}$ for TDT and $E[P_0(i)] \leq P_{avg}$ for FDT, where $E[\cdot]$ denotes the time average.

Each EHU has a battery that stores the harvested RF energy with a power conversion efficiency coefficient $\eta$, where $0 < \eta < 1$. Thus, the amount of power harvested into the battery of EHU $n$ at time slot $i$ is given by
\begin{equation} \label{rav15a}
P_{{in},n}(i) =\left\{
\begin{array}{rl}
a_i \eta  N_0 P_0(i) x_n(i), & \text{for TDT}
\\
\eta N_0 P_0(i) y_n(i), & \text{for FDT}.
\end{array}
\right.
\end{equation}

Let $B_n(i-1)$ denote the available power in the battery of EHU $n$ at the beginning of slot $i$, and let the battery's storage capacity be unlimited, i.e., $B_{max} \to \infty$. However, due to the inefficiency of the power amplifier (PA) of the EHU, the power transmitted by the EHU is assumed to be $1/\varepsilon$ times the power actually extracted (consumed) from its battery, where $\varepsilon$ is the PA efficiency coefficient ($\varepsilon > 1$). Therefore, when EHU $n$ transmits a codeword in slot $i$, the battery can supply at most a transmit power of $B_n(i-1)/\varepsilon$ in that slot. As a result, the desired transmit power of the codeword, $P_{d,n}(i)$, may not be available in the EHU's battery, in which case, the codeword is transmitted with some lower power, $P_{out,n}(i)$, i.e., $P_{out,n}(i) \leq P_{d,n}(i)$. Actually, the codeword is transmitted with power
\vspace{-1mm}
\begin{equation} \label{rav15b}
P_{{out},n}(i) =\left\{
\begin{array}{rl}
(1-a_i) \ P_{{out},n}'(i), & \text{for TDT}
\\
 P_{{out},n}'(i), & \text{for FDT}
\end{array}
\right.
\end{equation}
where
\vspace{-1mm}
\begin{equation} \label{rav15d}
P_{{out},n}'(i) = \min \left\{\frac{B_n(i-1)}{\varepsilon}, P_{{d},n}(i) \right\},
\end{equation}
such that the power extracted from the  battery is $\varepsilon P_{out,n}(i)$. At the end of slot $i$, the power remaining in the battery is
\begin{eqnarray}\label{rav15c}
    B_n(i)=B_n(i-1)+P_{{in},n}(i)- \varepsilon P_{out,n}(i) .
\end{eqnarray}

\vspace{-6mm}

\section{Optimal power and rate allocations}
\vspace{-0mm}
The capacity region of a standard MAC with perfect CSI available at all the nodes is defined by \cite[Eqs. (4)-(5)]{lit1}. Since the EHUs send their information simultaneously, the capacity region of the EH MAC, $C_{Region}$, is defined similarly as that of the standard MAC only by replacing the power variables in \cite[Eq. (4)]{lit1} with the EHUs' transmit powers $P_{out,n}(i)$, yielding
\begin{eqnarray} \label{rav2}
&&\hspace{-5mm}C_{Region}  = \Big \{ \bigcup \left(\bar R_1, \bar R_2, ..., \bar R_N \right):  \sum_{j \in S} \bar R_j \notag \\
&&\leq E \bigg [ \log_2\bigg(1 + \sum_{j \in \mathcal{S}} P_{out,j}(i) \, x_j(i)\bigg) \bigg ] \bigg \},\quad
\end{eqnarray}
where $\mathcal{S}$ is any subset of $\{1, 2, 3, ..., N\}$ and $\bar R_n$ is the $n$th EHU's information rate. Given an arbitrary set $P_{{out},n}(i), {\forall n,i}$, any set of rates $\bar R_n, {\forall n}$, which satisfies (\ref{rav2}) for all possible subsets $\mathcal{S}$, belongs to $C_{Region}$. In order to obtain the boundary surface of this capacity region, we follow the method presented in \cite[Lemma 3.10]{lit1}. We define a multidimensional plane $\sum_{n=1}^N \mu_n \bar R_n$, where $\mu_n, \forall n,$ are an arbitrary set of constants that satisfy $\sum_{n=1}^N\mu_n=1$. Due to the convexity of the capacity region (\ref{rav2}), for given $\mu_n, \forall n$, one point of the boundary surface of $C_{Region}$ is found as the solution of the following optimization problem
\begin{eqnarray} \label{novrav1}
\underset{\bar R_n,  P_{out,n}(i)} {\text{maximize}} && \sum_{n=1}^N \mu_n \bar R_n,  \quad s.t.\;\; \bar R_n \in C_{Region}, \forall n.
\end{eqnarray}
The entire boundary surface of $C_{Region}$ is obtained by solving (\ref{novrav1}) for all possible $\mu_n, \forall n$, such that $\sum_{n=1}^N\mu_n=1$.

Without loss of generality, let us assume that $\mu_1 \geq \mu_2 \geq ... \geq \mu_N$. Then, the solution of (\ref{novrav1}) over the variable $\bar R_n$, $\forall n$, is given by \cite[Eq. (16)]{lit1}, \cite[Eq. (6)]{lit2}
\begin{equation} \label{rav9}
\bar R_n^* = E \bigg[ \log_2 \left(1 + \frac{P_{{out},n}(i) x_n(i)}{1 + \sum_{k<n} P_{{out},k}(i)x_k(i)} \right) \bigg ],
\end{equation}
which is the optimal rate for an arbitrary $P_{out,n}(i), \forall n,i$. The achievability of the rate in (\ref{rav9}) is  as follows: Given $P_{{out},n}(i)$, $\forall n$, the EHU $n$ in time slot $i$ transmits a Gaussian distributed codeword (comprised of infinitely many symbols and spanning one time slot), which carries
\begin{equation} \label{novrav2}
R_n^*(i)= \log_2 \left(1 +  \frac{P_{{out},n}(i) x_n(i)}{1 + \sum_{k<n} P_{{out},k}(i)x_k(i)}\right)
\end{equation}
bits/symb. The BS on the other hand, decodes the EHUs' codewords in the order $N$, $N-1$, ..., $1$ and uses successive interference cancelation of the decoded codewords, [\ref{lit1}], [\ref{lit2}].

In order to find the optimal powers $P^*_{out,n}(i)$, $\forall n,i$, (\ref{rav9}) has to be inserted into  (\ref{novrav1}), and then (\ref{novrav1}) solved for $P_{out,n}(i)$, $\forall n,i$. Since $P_{out,n}(i)$ is given by (\ref{rav15b}), optimizing over $P_{{out},n}(i)$ is equivalent to optimizing $(a_i, P_{d,n}(i), P_0(i))$ for TDT, and $(P_{d,n}(i), P_0(i))$ for FDT.

\vspace{-4mm}
\subsection{Time Division Information and Power Transfer}
\vspace{-1mm}
Combining (\ref{rav9}) and (\ref{rav15b}), the optimal rate of EHU $n$ can be expressed as
\begin{equation} \label{rav9a}
\bar R_n^* = \hspace{-0.5mm} E \bigg [ (1-a_i)\log_2\hspace{-1mm} \bigg(1 \hspace{-0.5mm}+\hspace{-0.5mm} \frac{P_{{out},n}'(i) x_n(i)}{1 \hspace{-0.5mm} +\hspace{-0.5mm}  \sum_{k<n} P_{{out},k}'(i)x_k(i)} \bigg) \bigg ],
\end{equation}
with $P_{{out},n}'(i)$ given by (\ref{rav15d}), whereas (\ref{novrav1}) becomes
\begin{equation} \label{rav16}
\underset{P_{d,n}(i), P_0(i), a_i } {\text{maximize}} \ \ \sum_{n=1}^N \mu_n \bar R_n^* \notag
\end{equation}
\vspace{-4mm}
\begin{eqnarray}
   &C1:& P_{{out},n}'(i) =  \min\left\{B_n(i-1)/\varepsilon,  P_{{d},n}(i) \right\} \notag \\
&C2:& B_n(i) = B_n(i-1) + a_i  \eta N_0 P_0(i) x_n(i) \notag\\
&&\qquad\quad - (1-a_i) \, \varepsilon P_{{out},n}'(i)  \notag  \\
&C3:& E [a_i P_0(i)]  \leq P_{avg} \notag \\
&C4:& 0 \leq P_0(i) \leq P_{max},  \notag \\
&C5:& a_i \in \{0, 1\},
\end{eqnarray}
with $\bar R_n^*(i)$ given by (\ref{rav9a}). In (\ref{rav16}), $C1$ and $C2$ are actually the equalities (\ref{rav15d}) and (\ref{rav15c}), respectively, whereas $C3$-$C5$ represent the constrains on $a_i$ and $P_{0}(i)$. For finite $M$, solving (\ref{rav16}) is very difficult and may require non-causal CSI knowledge. However, for $M\to\infty$, optimization problem (\ref{rav16}) can be  simplified significantly, as shown in the following lemma.
\begin{lemma}
For $M\to\infty$, (\ref{rav16}) can be equivalently written as
\vspace{-1mm}
\begin{align}
\underset{P_{d,n}(i), P_0(i), a_i} {\text{maximize }} &\;\; \frac{1}{M} \sum_{i=1}^M \sum_{n=1}^N \mu_n \, R_{d,n}^*(i) \notag
\end{align}
\vspace{-4mm}
\begin{eqnarray} \label{rav17}
 \text{s.t.} \hspace{-5mm}&& \overline {C1}:  \frac{1}{M}\sum_{i=1}^M (1-a_i) P_{d,n}(i)   = \frac{1}{M}  \sum_{i=1}^M \eta' \, a_i P_0(i) x_n(i)  \notag \\
&& C3, \text{ } C4, \text{ and } C5 \text{ as in } (\ref{rav16}),
\end{eqnarray}
where $\eta' = \eta N_0/\varepsilon$ and
\vspace{-0mm}
\begin{equation} \label{rav17a}
R_{d,n}^*(i) = (1-a_i) \log_2 \left(1+\frac{ P_{d,n}(i) x_n(i)}{1 + \sum_{k<n} P_{d,k}(i)x_k(i)} \right) .
\end{equation}
\end{lemma}
\begin{IEEEproof}
Let us assume that $P_{in,n}(i)$ is a stationary random process, which means that $a_i$ and $P_0(i)$ are both stationary random processes (cf. Theorem 1). In this case, it was proven in [\ref{lit5}] that, when $M\to\infty$ and $B_{max}\to\infty$, the average rate of EHU $n$ in the EH MAC is maximized when the average desired power at its output, $E[(1-a_i) \varepsilon P_{{d},n}(i)]$, is set equal to its average harvested power, $\bar P_{{in},n}$. Since $\bar P_{{in},n} = E[\eta' \, a_i P_0(i) x_n(i)]$, the average rate of the EHU $n$ is maximized if, for $M\to\infty$,
\begin{eqnarray}{\label{test_2a}}
  \frac{1}{M}\sum_{i=1}^M  (1-a_i) \varepsilon P_{{d},n}(i)= \frac{1}{M}  \sum_{i=1}^M \eta N_0 \, a_i P_0(i) x_n(i).
\end{eqnarray}
 Moreover, it was proven in [\ref{lit5}] that, if (\ref{test_2a}) is satisfied, then,  the number of time slots in which $P_{{out},n}'(i) =\min\{ B_n(i-1)/\varepsilon,   P_{{d},n}(i)\}= B_n(i-1)/\varepsilon$ occurs  is negligible compared to the number of time slots in which $P_{{out},n}'(i) =\min\{ B_n(i-1)/\varepsilon, P_{{d},n}(i)\}=   P_{{d},n}(i)$  occurs. Furthermore, the time slots in which $P_{{out},n}'(i) = B_n(i-1)/\varepsilon$ occurs have negligible influence on the average information rate compared to the  time slots in which $P_{{out},n}'(i) =  P_{{d},n}(i)$ occurs. Hence, without any loss in the average rate, we can assume that $P_{{out},n}'(i) = P_{{d},n}(i)$ holds in all time slots. Inserting (\ref{test_2a}) and $P_{{out},n}'(i) = P_{{d},n}(i)$ into (\ref{rav16}), we obtain that constraint $C2$ is now not needed and can be removed, and thereby obtain (\ref{rav17}).
\end{IEEEproof}

The solution of  (\ref{rav17}) is given in the following theorem.
\begin{theorem}
The optimal "selection" rule is: $a_i^* = 0$ if
\begin{align} \label{rav23}
&\sum_{n=1}^N \left(\mu_n - \mu_{n+1}\right) \log_2 \left(1 + \sum_{k=1}^n P_{d,k}(i) x_k(i) \right)  \notag \\
&-\sum_{n=1}^{N} \lambda_n P_{d,n}(i) \geq P_0(i) \left( \eta' \sum_{n=1}^N \lambda_n x_n(i) - \lambda_0 \right),
\end{align}
and $a_i^* = 1$, otherwise. Optimal power allocation at the BS is
\begin{equation} \label{rav22}
P_0^{*}(i) =\left\{
\begin{array}{ll}
P_{max}, & \sum_{n=1}^N\lambda_n x_n(i) \geq \lambda_0/\eta' \ \text{and} \ a_i = 1
\\
\ \ 0, & \text{otherwise}.
\end{array}
\right.
\end{equation}
The optimal power allocation of EHU $n$ is $P_{d,n}^*(i) = (1 - a_i^*) P_{n}^*(i)$, where $P_{n}^*(i)$, $\forall n$ are given by [\ref{lit2}, Eq. (10)]. Specifically, when $a_i^* = 0$, EHU $n$ can either transmit or not depending on vector $\mathbf{x}(i)=[x_1(i),x_2(i),...,x_N(i)]$. The space of all possible vectors $\mathbf{x}(i)$ is divided into $2^N$ disjoint regions, such that the $j$th region is associated with the binary expansion $\{j_1, j_2,$ $..., j_k, ..., j_N \}$, where $j_k=1$ implies an active user and $j_k = 0$ implies a silent user. If the indices $n_1 < n_2 < \cdots <n_m< \cdots < n_l$ denote the positions of 1s in this binary expansion, then $N-l$ EHUs are silent and $l$ EHUs transmit with optimal powers [\ref{lit2}, Eq. (10)]
\begin{align} \label{rav6}
P_{n_1}^*(i) & =  \frac{\mu_{n_1}-\mu_{n_2}}{\lambda_{n_1}-\lambda_{n_2}\,x_{n_1}(i)/x_{n_2}(i)} - \frac{1}{x_{n_1}(i)} \notag \\
: & \notag \\
P_{n_m}^*(i) & =  \frac{\mu_{n_m}-\mu_{n_{(m+1)}}}{\lambda_{n_m} - \lambda_{n_{(m+1)}}\,x_{n_m}(i)/x_{n_{(m+1)}}(i)} \notag \\
&- \frac{\mu_{n_{(m-1)}} - \mu_{n_m}}{-\lambda_{n_m} + \lambda_{n_{(m-1)}}\,x_{n_m}(i)/x_{n_{(m-1)}}(i)} \notag \\
: & \notag \\
P_{n_l}^*(i) & = \frac{\mu_{n_l}}{\lambda_{n_l}} - \frac{\mu_{n_{(l-1)}} - \mu_{n_l}}{-\lambda_{n_l} + \lambda_{n_{(l-1)}}\,x_{n_l}(i)/x_{n_{(l-1)}}(i)} \notag \\
P_s^*(i) & = 0, \,\, \forall s \notin \{n_1,..., n_m, ..., n_l\} ,
\end{align}
Constants $\{\lambda_n\}_{n=1}^N$ and $\lambda_0$ are Lagrangian multipliers, which are determined from $\overline {C1}$ and $C3$ with ``$\leq$" replaced by ``$=$".
\end{theorem}

\vspace{-0mm}

\begin{IEEEproof}
Although (\ref{rav17}) is a non-convex optimization problem, we can still apply the Lagrange duality method to solve it, because it satisfies the \textit{time-sharing} condition in [\ref{lit3}] and has zero duality gap. To see this, we transform the inner sum of the cost function of (\ref{rav17})  to
\begin{equation} \label{rav18}
(1 - a_i) \sum_{n=1}^N \left(\mu_n - \mu_{n+1}\right) \log_2 \left(1 + \sum_{k=1}^n P_{d,k}(i) x_k(i) \right)
\end{equation}
where $\mu_1 \geq \mu_2 \geq ... \geq \mu_N$, and $\mu_{N+1} = 0$, and also note that $E[(1 - a_i^*) P_{d,n}^*(i)]$ and $[a_i^* P_0^*(i)]$ are increasing functions of $P_{avg}$. Thus, similarly to [\ref{lit4}, (P4)], we can conclude that the maximum value of optimization problem (\ref{rav17}) is concave in $P_{avg}$, and thus (\ref{rav17}) has zero duality gap.

In order to apply the Lagrange duality method, the Boolean constraint $C5$ is relaxed to a linear constraint, $\overline {C5}$: $0 \leq a_i \leq 1, \forall i$. In the following, we show that the optimal solution is satisfied at the boundaries of $\overline {C5}$, thus exactly satisfying $C5$. The Lagrangian is written as
\begin{align} \label{rav19}
&L = \frac{1}{M} \sum_{i=1}^M \sum_{n=1}^N \left(\mu_n - \mu_{n+1}\right) \left(1-a_i\right) \notag \\
&\times \log_2 \left(1 + \sum_{k=1}^n P_{d,k}(i) x_k(i) \right) - \frac{1}{M} \sum_{i=1}^M \sum_{n=1}^N \lambda_n (1 - a_i) P_{d,n}(i) \notag \\
&+ \frac{1}{M} \sum_{i=1}^M \eta' a_i P_0(i) \sum_{n=1}^N \lambda_n x_n(i) - \lambda_0 a_i P_0(i) + \nu_{1i}P_0(i) \notag \\
&+ \nu_{2i} [P_{max} - P_0(i)] + \tau_{1i} a_i + \tau_{2i} (1 - a_i),
\end{align}
where $\{\lambda_n\}_{n=1}^N$ and $\lambda_0$ are the non-negative Lagrange multipliers representing $\overline {C5}$ and $C3$, respectively, $\nu_{1i}$ and $\nu_{2i}$ correspond to $C4$, whereas $\tau_{1i}$ and $\tau_{2i}$ correspond to $\overline {C5}$.

By differentiating (\ref{rav19}) with respect to (w.r.t.) $P_0(i)$ and $a_i$, and then setting both derivatives to zero, we obtain
\begin{equation} \label{rav20}
\frac{dL}{dP_0(i)} = a_i \left(\eta' \sum_{n=1}^N \lambda_n x_n(i) - \lambda_0 + \nu_{1i} - \nu_{2i} \right) = 0
\end{equation}
\vspace{-3mm}
\begin{align} \label{rav21}
\frac{dL}{da_i} & = -\sum_{n=1}^N \left(\mu_n - \mu_{n+1}\right) \log_2 \left(1 + \sum_{k=1}^n P_{d,k}(i) x_k(i) \right) \notag \\
& + \sum_{n=1}^N \lambda_n P_{d,n}(i) + \eta' P_0(i) \sum_{n=1}^N \lambda_n x_n(i) \notag \\
& - \lambda_0 P_0(i) + \tau_{1i} - \tau_{2i} = 0.
\end{align}
According to the Karush-Kuhn-Tucker (KKT) conditions, complimentary slackness should be satisfied, $\forall i$: $\nu_{1i}P_0(i) = \nu_{2i} [P_{max} - P_0(i)] = \tau_{1i} a_i = \tau_{2i} (1 - a_i) = 0$. Let us assume that $0 < P_0(i) < P_{max}$ and $0 < a_i < 1$ hold. Then, according to the KKT conditions, the following must hold  $\nu_{1i} = \nu_{2i} = \tau_{1i} = \tau_{2i} = 0$. However, by setting $0 < P_0(i) < P_{max}$, $0 < a_i < 1$, and $\nu_{1i} = \nu_{2i} = \tau_{1i} = \tau_{2i} = 0$ in both (\ref{rav20}) and (\ref{rav21}), we see that  (\ref{rav20}) and (\ref{rav21}) cannon hold for an arbitrary $i$, due to  $x_n(i)$ being random. Therefore, $P_0(i) \in \{0, P_{max}\}, \forall i$, yielding (\ref{rav22}), and $a_i \in \{0, 1\}, \forall i$, yielding (\ref{rav23}).

We now differentiate (\ref{rav19}) w.r.t. $P_{d,n}(i), \forall n$, set them to zero, and thus obtain a set of $N$ equations. For uplink transmission ($a_i = 0$), this set is the same as [{\ref{lit2}}, Eq. (9)]. Thus, $P_{d,n}^*(i) = (1 - a_i) P_{n}^*(i)$, where $P_{n}^*(i)$ is given by (\ref{rav6}).
\end{IEEEproof}

The rates $R_{d,n}(i)$ selected according to (\ref{rav17a}), with $P_{d,n}(i)$ and $a_i$ selected according to (\ref{rav6}) and (\ref{rav23}), respectively, yield to the boundary of the achievable EH MAC rate region.

\vspace{-3mm}

%%%%%%%%%%%%%%%%%%%%%%%%%%%%%%%%%%%%%%%%%%
\subsection{Frequency Division Information and Power Transfer}
We  again  consider $M\to\infty$ slots. In order to maximize the achievable rate region of the EH MAC, we need to solve a similar optimization problem as in (\ref{rav16}), where the ``scheduling" variables $a_i$ and $(1-a_i)$ are omitted and $P_{{out},n}(i) = P_{{out},n}'(i)$,
\begin{equation}
\underset{P_{d,n}(i), P_0(i)} {\text{maximize}} \ \ \sum_{n=1}^N \mu_n \bar R_n^* \notag
\end{equation}
\vspace{-5mm}
\begin{eqnarray} \label{rav11}
\text{s.t.} &C1:& P_{{out},n}(i) = \min\left\{B_n(i-1)/\varepsilon, P_{d,n}(i) \right\} \notag \\
&C2:& B_n(i) = B_n(i-1) + \eta N_0 P_0(i) y_n(i) \notag \\
&&- \varepsilon P_{out,n}(i) \notag \\
&C3:& E \left[P_0(i) \right] \leq P_{avg} \notag \\
&C4:& 0 \leq P_0(i) \leq P_{max},
\end{eqnarray}
where the optimal rate of EHU $n$, $\bar R_n^*$, is given by (\ref{rav9}). When $M \to \infty$, the equivalent optimization problem is analogous to (\ref{rav17}), (\ref{rav17a}), with variables $a_i$ and $(1-a_i)$ omitted. Using a similar approach as before, the EHUs' optimal power allocations are found as $P_{d,n}^{*}(i) = P_n^*(i)$, with $P_n^*(i)$ given by (\ref{rav6}), whereas the optimal powers at the BS are found as
\begin{equation} \label{rav14}
P_0^{*}(i) =\left\{
\begin{array}{ll}
P_{max}, & \sum_{n=1}^N\lambda_n y_n(i) \geq \lambda_0/\eta'
\\
\ \ 0, & \text{otherwise} .
\end{array}
\right.
\end{equation}

\noindent The proof is omitted for brevity.

\vspace{-3mm}

\section{Numerical Results}
We illustrate our results for the case of a multiple-access system with two EHUs ($N = 2$) and a BS. Since Rayleigh fading is considered, $x_n(i)$ and $y_n(i)$ follow an exponential distribution, with $E[x_n(i)]$ and $E[y_n(i)]$ being inversely proportional to the deterministic path loss (PL). The PLs of the uplink and downlink channels are assumed to be identical for both EHUs and for both FDT and TDT, i.e., $PL_X = PL_Y = 10^5$ and $E[x_1(i)] = E[y_1(i)] = E[x_2(i)] = E[y_2(i)] = 1/(N_0 \, PL_X)$.

The average and the maximum output powers of the BS are $P_{avg} = 10$ and $P_{max} =  50$, respectively. We set $N_0 = 10^{-5}$, $\eta = 0.5$, $\varepsilon = 5$, such that $\eta' = 10^{-6}$. Because the average power harvested by each EHU is very low (less than $\eta' P_{max} E[y_n(i)]$), the storage capacity of an EH battery is practically infinite with today's technology, i.e., $B_{max} \to \infty$.

Fig. 1 depicts the achievable rate region of the EH MAC for FDT (full lines) or TDT (dashed lines) systems. Various numbers of transmission slots ($M$) are considered, including $M\to\infty$. For a given $M$, the power allocations of the BS and EHUs are calculated according to the respective optimal power allocations for $M\to\infty$, derived in Section III. For $N=2$, (\ref{rav6}) reduces to [\ref{lit2}, Eqs. (11)-(12)].

Each rate pair $(\bar R_1, \bar R_2)$ on the given curve corresponds of the priority coefficient pair $(\mu_1, \mu_2)$, where $\mu_1 + \mu_2 = 1$. The rate pairs are calculated according to $\bar R_1 = (1/M) \sum_{i=1}^M R_{1}(i)$ and $\bar R_2 = (1/M) \sum_{i=1}^M R_{2}(i)$, respectively, where $R_n(i)$ is calculated from (\ref{novrav2}). An infinite $M$ achieves the boundary of the achievable EH MAC rate region (red lines), whereas lower rates are achieved for smaller $M$ (blue lines). For finite $M$, the proposed online solution is simple but suboptimal, whereas it is optimal for $M\to\infty$. We note that,  for $M\to\infty$  the boundary surface of the EH MAC with FDT is identical to the boundary surface of the standard (non-EH) MAC with average power constraint $\eta' P_{avg} E[y_n(i)]$.

Due to the severe attenuation of the wireless power transfer, both FDT and TDT schemes allocate very few time slots for uplink data transmissions, thus achieving comparatively low rates. Although FDT transfers power and information simultaneously, for TDT only few time slots are lost for wireless power transfer compared to FDT. Thus, FDT outperforms TDT only by a small margin.

\vspace{-3mm}

\begin{figure}[tbp]
\centering
\includegraphics[width=3.5in]{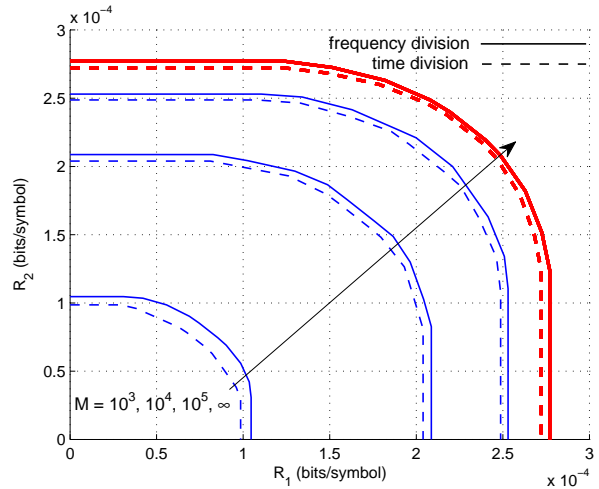} \vspace{-7mm}
\caption{Achievable rate region for a system with two EHUs, where $\eta = 0.5, \varepsilon = 5, N_0 = 10^{-5}, P_{avg} = 10$, and $P_{max} = 50$} \vspace{-4mm}
\label{fig1}
\end{figure}

\end{document}